\documentclass[onecolumn,preprintnumbers,amsmath,amssymb]{revtex4}
\usepackage{graphicx}
\usepackage{dcolumn}
\usepackage{bm}

\begin{document}

\title{Exact eigenfunctions of $N$-body system with quadratic pair potential}

\author{Zhaoliang Wang}
 \email{wzlcxl@mail.ustc.edu.cn; Tel./fax:+865513607061}
\author{Anmin Wang}
 \email{anmwang@ustc.edu.cn}
\author{Yang Yang}
\author{Xuechao Li}
\affiliation{Department of Modern Physics, University of Science and
Technology of China, Hefei, Anhui, China.}
\begin{abstract}
We obtain all the exact eigenvalues and the corresponding eigenfunctions of N-body Bose and Fermi systems with Quadratic Pair Potentials in one dimension. The original first excited state or energy level is disappeared in one dimension, which results from the operation of symmetry or antisymmetry of identical particles. In two and higher dimensions, we give all the eigenvalues and the analytical ground state wave functions and the degree of degeneracy. By comparison, we refine Avinash Khare's results by making some items in his article precisely.
\end{abstract}



\maketitle

\section{Introduction}
\label{sec:level1}

Operating on quantum many-body systems provide a way to understand the world, so hunting an exactly solved quantum many-body model becomes more important, especially for pair interaction models. Over a period of more than seventy years, there has been much successes in dealing with quantum many-body problems, but only a few models can be solved exactly. Calogero-Sutherland (CS) model \cite{Calogero, Sutherland1, Sutherland2, Barnali, Levy} is one celebrated example of solvable many-body problems. It has been wide applications in quantum chaos and fractional statistics \cite{Simons}. As a special case of CS model, One dimensional system with quadratic pair potentials was studied by H. R. Post \cite{Post}, who obtained the ground state energy. J. M. Levy-Leblond \cite{Levy} obtained the energy spectrum in one dimension and the ground state energy in three dimension. Zhong-Qi Ma \cite{Ma} considered three particle system and he pointed out that some states were disappeared by antisymmetric operations. None of them have given the exact eigenfunctions and as we know that eigenfunctions are important for us to compute many kinds of correlations, which motivate us to restudy this system and obtain the exact eigenfunctions.

 Up to now, Calogero-Sutherland model is still not a completely solved problem. Calogero \cite{Calogero, Calogero2} and Khare \cite{Ray, Khare} made a big step forward that they gave partial exact solutions of the Calogero-Sutherland model, such as the Boson and Fermion ground state and radial excitations over it. In this article, we will first exhibit the exact eigenvalues and the corresponding eigenfunctions of N-body Bose and Fermi systems with Quadratic Pair Potentials in one dimension. Second, we will give our findings of Fermi system with quadratic pair potentials in two and higher dimensions. At last, our results will be compared with the existing results of Khare.

\section{$N$-body System with Quadratic Pair Potentials in one dimension}\label{sec:level2}

The Hamiltonian of one dimensional $N$-body problems with quadratic pair potentials is
\begin{equation}\label{1}
  H=\sum^N_{i=1}-\frac{\hbar^2}{2m}\frac{\partial^2}{\partial x_i^2}+\sum^N_{i<j}\frac{1}{2}m\omega^2\left(x_i-x_j\right)^2.
\end{equation}

 With the Jacobin coordinates
\begin{equation}\label{3}
  \xi_i=\begin{cases}\frac{1}{i}\sum_{j=1}^{i}x_j-x_{i+1}, & \text{$1\leqslant i\leqslant N-1$},\\
  \frac{1}{N}\left(x_1+x_2+\cdots+x_{_N}\right), &\text{$i=N$},
  \end{cases}
\end{equation}
 where $\xi_i$ is the distance between the center of mass of the first $i$ particles and the $(i+1)th$ particle, we have
\begin{align}\label{5}
  \begin{split}
  &H\psi=(H_c+H_\xi)\psi=E\psi,\\
  &H_c=-\frac{\hbar^2}{2m}\frac{\partial^2}{\partial \xi_{_N}^2},\qquad H_\xi=\sum^{N-1}_{i=1}H_i,\qquad
  H_i=\left[-\frac{\hbar^2}{2\mu_i}\frac{\partial^2}{\partial \xi^2_i}+\frac{1}{2}\mu_i\left(\sqrt{N}\omega\right)^2 \xi_i^2\right].
  \end{split}
\end{align}
with the reduced mass $\mu_i=\frac{i}{i+1}m$. From equations (\ref{5}) we know that although any two particles interact with each other with a vibration frequency $\omega$, it can be decomposed into multiple quasi-particle's vibration with a different frequency $\sqrt{N}\omega$. without considering the centroid movement, the solution of equations (\ref{5}) is
\begin{align}
  &E_\xi=\sum_{i=1}^{N-1}E_{n_i}=\left(n_i+\frac{1}{2}\right)\hbar\omega',\\
  &\psi_\xi=\prod^{N-1}_{i=1}\varphi_{n_i}\left({\xi_i}\right)=\prod^{N-1}_{i=1}N_{n_i,i}\exp\left(-\frac{1}{2}\alpha_i^2\xi_i^2\right)H_{n_i}\left(\alpha_i\xi_i\right)
\end{align}
where
\begin{equation*}
  N_{n_i,i}=\left[\frac{\alpha_i}{2^{n_i}n_i!\sqrt{\pi}}\right]^\frac{1}{2},\,\,\,\, \alpha_i=\sqrt{\frac{\mu_i\omega'}{\hbar}}\,\, ,\,\,\,\, \omega'=\sqrt{N}\omega
\end{equation*}

Taking the first $i$ particles as a system, then $H_i$ can be regarded as the newly emerged Hamiltonian while adding another one particle to the system and it is independent with the system constituted by the first $i$ particles.

As we all know, Bose system should be exchange symmetric, while Fermi system should be exchange anti-symmetric. Because of the interactions, it is not easy to get the eigenfunctions of the identical particle systems. The eigenfunctions can be written as
\begin{equation}\label{eqn6}
    \psi=\sum_P a(P)P\varphi_{n_1}\left(\xi_1\right)\varphi_{n_2}\left(\xi_2\right)\cdots\varphi_{n_{_{N-1}}}\left(\xi_{_{N-1}}\right),
\end{equation}
where $P$ is an exchange operator of any two particles or an permutation of all particles. For Bose system, $a(P)=1$; While for Fermi system,
\begin{equation}
  a(P)=\begin{cases}
  1, & \text{$P$ is an even permutation},\\
  -1, & \text{$P$ is an odd permutation}.
  \end{cases}
\end{equation}
Take two-particle Fermi system as an example,
\begin{equation}\label{eqn11}
    \psi=\varphi_{n_1}(\xi_1)-P_{12}\varphi_{n_1}(\xi_1)=\varphi_{n_1}(x_1-x_2)-\varphi_{n_1}(x_2-x_1)
                =\begin{cases}
                2\varphi_{n_1}(\xi_1), & \text{$n_1$ is odd},\\
                0, & \text{$n_1$ is even}.
                \end{cases}
\end{equation}

 We will use the subscript $n_i$ as the quantum number to constitute the sets $\{n_1,n_2,\cdots,n_{_{N-1}}\}$ to represent the eigenfunction. By calculation, we found such an ansatz:

{\em All the states can be obtained by only changing the last quantum number $n_{_{N-1}}$ when the system constituted by the first $N-1$ particles is in ground state.\/}

I have to declare that every quantum number plays a role in solving the states. This ansatz, the correctness of which can be tested and verified by substituting the final results into equation (\ref{1}), is just one way to obtain all the states.

For Bose system, it is clear that the combination $n_1=n_2=\cdots=n_{_{N-2}}=0$ is the ground state of the system constituted by the first $N-1$ particles, so equation (\ref{eqn6}) changes to
\begin{equation}
  \psi^B=\sum_{i=1}^N P_{iN}\varphi_0\left(\xi_1\right)\cdots\varphi_0\left(\xi_{_{N-2}}\right)\varphi_{n_{_{N-1}}}\left(\xi_{_{N-1}}\right)
\end{equation}

By calculation, we found the energy spectrum and the corresponding ground state and excited state eigenfunctions,
\begin{align}
  &E_k^B=\begin{cases}
  \frac{N-1}{2}\sqrt{N}\hbar\omega, & \text{$k=0$},\\
  E_0+(k+1)\sqrt{N}\hbar\omega, & \text{$k\neq0$}.
  \end{cases}\\
  &\psi_0^B=C_0\exp\left(-\frac{m\omega}{2\sqrt{N}\hbar}\sum_{i>j}^{N-1}x_{ij}^2\right),\qquad \qquad
\end{align}
\begin{equation}
    \psi_k^B=C_k\exp\left(-\frac{m\omega}{2\sqrt{N}\hbar}\sum_{i>j}^{N-1}x_{ij}^2\right)\sum_{j=1}^NH_{k+1}\left[\sqrt{\frac{m\omega}{\sqrt{N}(N-1)\hbar}}\left(\sum_{i=1}^Nx_i-Nx_j\right)\right],
\end{equation}
where$x_{ij}=x_i-x_j$ and $H_{k+1}$ is the Hermite polynomials. Note that the original existed first excited state or energy level is disappeared and all the states are non-degenerate.

In the following, we will give the solutions of Fermi system briefly. With the ansatz and the principle of mathematical induction, we can demonstrate that $\{1,2,3,\cdots,N-1\}$ is the ground state wave function of the system.
\begin{equation}\label{12}
    \psi_0^F=\sum_P a(P)P\varphi_1\left(\xi_1\right)\varphi_2\left(\xi_2\right)\cdots\varphi_{_{N-1}}\left(\xi_{_{N-1}}\right)=C_0\prod_{i>j}^Nx_{ij}\exp\left(-\frac{m\omega}{2\sqrt{N}\hbar}\sum_{i>j}^{N-1}x_{ij}^2\right)
\end{equation}
  where $C_0$ is the normalization constant, satisfying $\int_{-\infty}^\infty\left|\psi_0\right|^2 \,d\xi_1d\xi_2\cdots d\xi_{_{N-1}}=1$, then
\begin{equation*}
  C_0=\left[\left(\frac{m\omega}{\hbar}\right)^{\frac{N^2-1}{2}}\pi^{\frac{1-N}{2}}N^{\frac{N^2-3}{4}}\prod_{i=1}^N \frac{2^{i-1}}{i!}\right]^{1/2}.
\end{equation*}

 Here we have to clarify that the set $\{1,2,3,\cdots,N-1\}$ is one kind of combinations for the ground-state wave function, but it is not the only one kind that stands for the ground-state wave function. For three-particle system as an example, $\{1,2\}$ and $\{3,0\}$ are both combinations for the ground-state and both of which give the same result as expressed in equation (\ref{12}) when $N=3$. Therefore, it is non-degenerate for the ground-state of three-particle system. From our calculation we obtain that all states of the system are non-degenerate in one dimension, and this result is against with that of J. M. Levy-Leblond \cite{Levy}. As all the eigenstates of this system are non-degenerate, we will not consider other possible combinations except $\{1,2,3,\cdots,N-1\}$ to represent the ground state. In the following, we calculate all the eigenstates by the mathematical induction.

 From equation (\ref{eqn11}), we know $\psi_0^{(2)}=2\varphi_1(\xi_1)$, which is just in accord with equation (\ref{12}). We suppose equation (\ref{12}) is right for $(N-1)$-particle system, then when the particle number of the system is $N$, define
\begin{equation}\label{13}
 \psi(\lambda)^F=\sum_{P} a(P)P\varphi_1\left(\xi_1\right)\cdots\varphi_{_{N-2}}\left(\xi_{_{N-2}}\right)\varphi_{_\lambda}\left(\xi_{_{N-1}}\right).
\end{equation}
From the above ansatz we know all the possible eigenfunctions of $N$-particle Fermi system could be obtained by solving equation (\ref{13}). To avoid loss of continuity, we have relegated the process to Appendix. The main results are as follows:\\
$(a)$ For $\lambda <N-1$, $\psi(\lambda)^F=0$; For $\lambda =N-1$, equation (\ref{12}) is right for the ground state of $N$-particle system;\\
$(b)$ For $\lambda =N$, $\psi(N)^F=0$. $\psi(N)^F$ should be the first excited state, but it is missing, just like the missing of the first excited state of Boson system. We have not manipulated the eigenvectors except for exchanging any two particles' coordinates. It is safe for us to say that exchange symmetry or anti-symmetry result in the loss of the original existed first excited state or energy level.\\
$(c)$ For $k=\lambda-N>0$, we could get all the excited states of the system. The $k$th excited state is $\{1,2,\cdots,N-2,N+k\}$, and its eigenfunction is
\begin{equation}\label{14}
    \psi_k^F=C_k\prod_{i>j}^Nx_{ij}\exp\left(-\frac{m\omega}{2\sqrt{N}\hbar}\sum_{i>j}^Nx_{ij}^2\right)
    \sum_{i=0}^{\frac{k+\epsilon(k)}{2}}\left[\sum_{j=0}^{2i+1-\epsilon(k)}\left(\frac{4N\sqrt{N}m\omega}{(N-1)\hbar}\right)^i
    \frac{(-1)^{i+j}\sigma_1^jV_{(2i-j+1-\epsilon(k))}}{(\frac{k+\epsilon(k)}{2}-i)!j!(N+2i-j-\epsilon(k))!N^j}\right]
\end{equation}
where $\sigma_i$ is the elementary symmetric polynomial. For $N$-variable polynomials,
\begin{equation}\label{15}
  \sigma_i=\begin{cases}
  1, & \text{$i=0$}\\
  \sum_{j_1<j_2<\cdots < j_i}^N x_{j_1}x_{j_2}\cdots x_{j_i}, & \text{$1 \leqslant i \leqslant N$},\\
  0, & \text{else}.
  \end{cases}
\end{equation}
\begin{equation}\label{16}
  V_i=\begin{cases}
  1, & \text{$i=0$},\\
  \begin{vmatrix}
  \sigma_1 & \sigma_2 & \sigma_3 & \cdots & \sigma_i\\
  \sigma_0 & \sigma_1 & \sigma_2 & \cdots & \sigma_{i-1}\\
  0        & \sigma_0 & \sigma_1 & \cdots & \sigma_{i-2}\\
  \vdots   & \vdots   & \vdots   & \ddots & \vdots\\
  0        & 0        & 0        & \cdots & \sigma_1
  \end{vmatrix}, & \text{$i\neq 0$}.
  \end{cases}\qquad \qquad
\end{equation}
\begin{equation}\label{17}
  \epsilon(k)=\begin{cases}
  0, & \text{$k$ is even},\\
  1, & \text{$k$ is odd}.
  \end{cases}\qquad \qquad \qquad \qquad
\end{equation}

The Fermi exchange anti-symmetry is reflected in the factor $\prod_{i<j}^Nx_{ij}$, and it is lucky for us to see this factor appears in all eigenfunctions. The rest factors that appear in $\psi_0$ and $\psi_k$ are all functions of the elementary exchange symmetric polynomials.

From $(c)$, we know that $\{1,2,3,\cdots,N-2,N+i\}$ is one kind of combinations standing for the $i$th excited states of $N$-particle system. The exact eigenvalues corresponding to $\psi_0$ and $\psi_k$ of the Fermi system with Quadratic Pair Potentials in one dimension are

\begin{equation}
   E_i=\begin{cases}
   \frac{1}{2}\left(N^2-1\right)\sqrt{N}\hbar\omega, & \text{$i=0$}\\
   E_0+(i+1)\hbar\sqrt{N}\omega, & \text{else}.
   \end{cases}
\end{equation}

\section{Solutions of D-dimensional Fermi system}\label{sec:level3}
 the Hamiltonian is
\begin{equation}\label{18}
  H_T=\sum^N_{i=1}-\frac{\hbar^2}{2m}\frac{\partial^2}{\partial\bm{r}_i^2}+\sum^N_{i<j}\frac{1}{2}m\omega^2\left(\bm{r}_i-\bm{r}_j\right)^2
   =H_x+H_y+\cdots+H_\eta,
\end{equation}
with $\bm{r}=x\overrightarrow{e_x}+y\overrightarrow{e_y}+\cdots+\eta\overrightarrow{e_\eta}$ the D-dimensional coordinate vector. For Bose system, the ground state is just like equation (11), except changing $x_{ij}$ to $\bm{r}_{ij}$. For Fermi system, things are a bit more complicated. we just obtain the ground state energy and eigenfunction through the summary of the calculated results about fixed $N$. The correctness of all the following equations can be verified by checking whether they meeting equation (\ref{1}) or not.
\begin{equation}\label{19}
  E_{0,f}^{(D)}=\hbar\omega\sqrt{N}\left[(K+\frac{D}{2})N-\frac{1}{(D+1)!}\prod_{i=0}^D(K+i)-\frac{D}{2}\right]
\end{equation}
where $K$ is the integer obeying
\begin{equation}\label{20}
  \frac{1}{D!}\prod_{i=0}^{D-1}(K+i)\leqslant N-1<\frac{1}{D!}\prod_{i=1}^D(K+i),
\end{equation}
the result of J. M. Levy-Leblond \cite{Levy} is a special case for $D=3$ here. For large $N$, the result is asymptotically changes to
\begin{equation}\label{21}
  E_{0,f}^{(D)}\xrightarrow{N\rightarrow\infty}\hbar\omega\sqrt{N}\left\{\frac{DN}{D+1}[D!(N-1)]^{1/D}+O(N)\right\}.
\end{equation}
this result is the same as equation (20) in Khare's article \cite{Khare}, except the last small amount term.

For the ground state, the behavior the these Fermions are just like non-interacting Fermions occupying the states of a fictitious Hamiltonian whose energy levels are divided into different $k$ series, $k=1,2,3,\cdots$, and every $k$ series has a suppositional degree of degeneracy of $\frac{1}{(D-1)!}\prod_{i=1}^{D-1}(K+i)$ (it is 1 in one dimension). These non-interacting Fermions will be first to fill in the low $k$ series energy levels, then continue to fill in the higher energy levels. While the number of Fermions $N$ is just filled any series to the full, the system is non-degenerate; While there are $n$ Fermions left, it equals to choose $n$ states from $\frac{1}{(D-1)!}\prod_{i=1}^{D-1}(K+i)$ states for the left $n$ Fermions to occupy, which causes the degeneracy of the realistic Hamiltonian. This fictitious degeneracy result in an interesting fact that, for large particle number Fermi system, the value of the ground state energy decreases while the dimension $D$ of the system increases. And the relation between the ground state energy, the dimension $D$ and the particle number $N$ is plotted in Figure 1.

\begin{figure}[h]
\includegraphics[scale=0.4, bb=8 14 787 576]{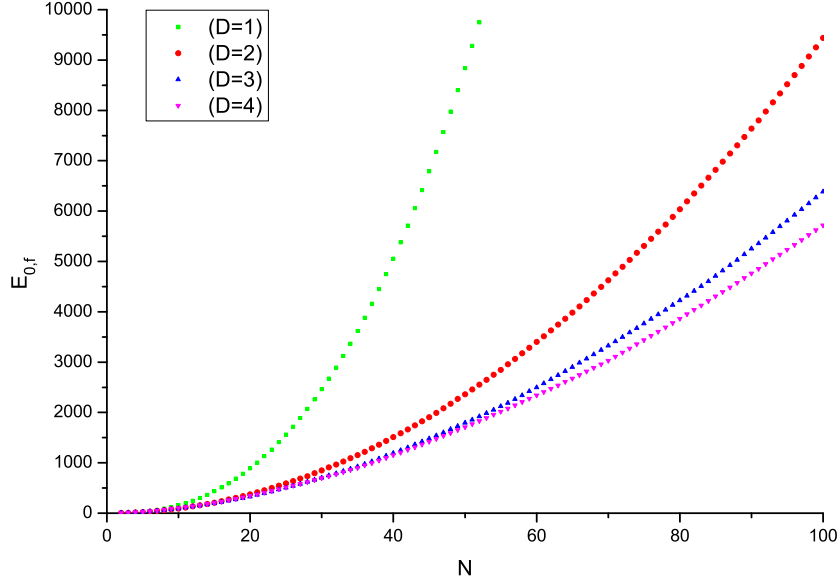}
\caption{(color online). Plots of the ground state energy $E_{0,f}$ with particle number $N$ in different dimensions. Here we set $\hbar\omega=1$}
\end{figure}

The corresponding ground state eigenfunction is
\begin{equation}\label{22}
  \Psi_{0,f}^{(D)}=C\psi_{_S}\exp\left(-\frac{m\omega}{2\sqrt{N}\hbar}\sum_{i>j}^{N-1}|\bm{r}_i-\bm{r}_j|^2\right),
\end{equation}
\begin{equation}\label{23}
  \psi_{_S}=\begin{vmatrix}
  L_0^1(1) & L_1^1(1) & \cdots & L_{K-1}^{d}(1) & L_K^{\mu_1}(1) & \cdots & L_K^{\mu_{N-n}}(1)\\
  L_0^1(2) & L_1^1(2) & \cdots & L_{K-1}^{d}(2) & L_K^{\mu_1}(2) & \cdots & L_K^{\mu_{N-n}}(2)\\
  \vdots & \vdots & \ddots & \vdots & \vdots & \ddots & \vdots\\
  L_0^1(n) & L_1^1(n) & \cdots & L_{K-1}^{d}(n) & L_K^{\mu_1}(n) & \cdots & L_K^{\mu_{N-n}}(n)\\
  L_0^1(n+1) & L_1^1(n+1) & \cdots & L_{K-1}^{d}(n+1) & L_K^{\mu_1}(n+1) & \cdots & L_K^{\mu_{N-n}}(n+1)\\
  \vdots & \vdots & \ddots & \vdots & \vdots & \ddots & \vdots\\
  L_0^1(N) & L_1^1(N) & \cdots & L_{K-1}^{d}(N) & L_K^{\mu_1}(N) & \cdots & L_K^{\mu_{N-n}}(N)
  \end{vmatrix},
\end{equation}
where
\begin{equation*}
  n=\frac{1}{D!}\prod_{i=0}^{D-1}(K+i),\quad d=\frac{1}{(D-1)!}\prod_{i=0}^{D-2}(K+i),
\end{equation*}
\begin{equation*}
  L_K^k(i)=x_i^{k_x}y_i^{k_y}\cdots \eta_i^{k_\eta},\quad 1\leqslant k\leqslant\frac{1}{(D-1)!}\prod_{i=1}^{D-1}(K+i).
\end{equation*}
Note that $k$ is just a serial number standing for the combination $\left\{k_x,k_y,\cdots,k_\eta\right\}$, which satisfies
\begin{equation*}
  k_x+k_y+\cdots+k_\eta=K.
\end{equation*}

In equation (\ref{23}), the first n lines and the first n columns stands for the first $K-1$ series being filled with Fermions. Then there are $N-n$ Fermions left, So the degeneracy of the ground state is
\begin{equation}\label{24}
  \binom{\frac{1}{(D-1)!}\prod_{i=1}^{D-1}(K+i)}{N-\frac{1}{D!}\prod_{i=0}^{D-1}(K+i)},
\end{equation}
where $\tbinom{n}{m}=\frac{n!}{m!(n-m)!}$ is the binomial coefficient.

For the ground state eigenfunctions of three Fermion system in three dimension as an example. From equation (\ref{20}), we know $K=1$. So the degeneracy is 3, and
\begin{equation}
  L_1^1(i)=x_i,\quad L_1^2(i)=y_i,\quad L_1^3(i)=z_i.
\end{equation}
There are two Fermions left to fill the $K=1$ series. Substitute any two of the three $L_1^k(i)$ into equation (\ref{23}), three $\psi_S$ will be obtained correspondingly.
\begin{align*}
  \psi_{S_1}^{N=3}&=x_2y_3-x_3y_2+x_1y_2-x_2y_1+x_3y_1-x_1y_3,\\
  \psi_{S_2}^{N=3}&=x_2z_3-x_3z_2+x_1z_2-x_2z_1+x_3z_1-x_1y_3,\\
  \psi_{S_3}^{N=3}&=y_2z_3-y_3z_2+y_1z_2-y_2z_1+y_3z_1-y_1z_3.
\end{align*}

From our calculation, we find some states are missing, but there are no energy levels missing because of the degeneracy. The exact eigenvalues of the system with Quadratic Pair Potentials in D-dimension are
\begin{equation}
  E_{i,f}^{(D)}=E_{0,f}^{(D)}+i\sqrt{N}\hbar\omega.
\end{equation}

Comparing the equation (\ref{22}) with the existing result (29) in Khare's article \cite{Khare}, we find that $\psi_{_S}$ in his article is just the $\psi_{_S}$ in this article and we make his energy exact because of our exact ground state energy. we rewrite his Hamiltonian and the $N$-Fermion ground state and radial excitations over it as follows.
\begin{align}
  H=&-\frac{1}{2}\sum_{i=1}^N\nabla_i^2+g\sum_{i<j}^N\frac{1}{\bm{r}_{ij}^2}+\frac{1}{2}\sum_{i<j}^N\bm{r}_{ij}^2
    +G\sum_{i<j,i\leqslant k,j\leqslant k}^N\frac{\bm{r}_{ki}\cdot\bm{r}_{kj}}{\bm{r}_{ki}^2\bm{r}_{kj}^2},\\
  \psi_n=&\prod_{i<j}^N\left|\bm{r}_i-\bm{r}_j\right|^{\Lambda_D^f}\psi_{_S}\exp\left(-\frac{1}{2\sqrt{N}}\sum \bm{r}_{ij}^2\right)
         \times L_n^{\Gamma_D^f-(D/2)} \left( \frac{1}{\sqrt{N}}\sum \bm{r}_{ij}^2\right),\\
  E_n^f=&E_{0,f}^{(D)}+\sqrt{N}\left[2n+\frac{1}{2}N(N-1)\Lambda_D^f\right].
\end{align}
Here we have set $\hbar=m=\omega=1$. Note that $E_{0,f}$ and $\psi_{_S}$ are exactly expressed by equation (\ref{19}) and (\ref{23}) respectively.

\section{conclusions}

We investigate a quantum system of which identical particles interact with each other with quadratic pair potentials. By calculations, we obtain the energy spectrum and the corresponding eigenfunctions which are non-degenerate and proved that the original first excited state or energy level disappears. In two and higher dimensions, we present the energy spectrum and the degree of the ground state and eigenfunctions. For large particle number Fermi system, the value of the ground state energy decreases while the dimension increases. Through the comparison with Avinash Khare's results, we give the exact expressions of $\psi_{_S}$ and $e_0^f$ of the Calogero model in his article. Besides these already solved problems, there are also other unsolved problems. For example, the angular excitations of the Calogero model. We will continue studying this problem later.

Acknowledgments: This work is supported by the National Natural Science Foundation of China under Grant No. 10975125.

\appendix

\section{Appendix}

To begin the demonstration, we exhibit a formula-the hyper-Vandermonde determinant. With the definition in equation (\ref{15}) and (\ref{16}), we could generalize the application range of the result in Xiaolin Wang's article \cite{xiaolin}.
\begin{multline}\label{A1}
   \sum_{k=1}^N a(P_{kN})P_{kN}\prod_{i>j}^{N-1}\left(x_i-x_j\right) x_N^\gamma
   =\begin{vmatrix}
    1 & 1 & 1 & \cdots & 1\\
    x_1 & x_2 & x_3 & \cdots & x_N\\
    x_1^2 & x_2^2 & x_3^2 & \cdots & x_N^2\\
    \vdots & \vdots & \vdots & \ddots & \vdots\\
    x_1^{N-2} & x_2^{N-2} & x_3^{N-2} &\cdots & x_{_N}^{N-2}\\
    x_1^\gamma & x_2^\gamma & x_3^\gamma & \cdots & x_{_N}^\gamma
    \end{vmatrix}\\
    =\begin{cases}
    0, & \text{$\gamma<N-1$},\\
    V_{_N}^{(\gamma-N+1)}\prod_{i>j}^{N-1}\left(x_i-x_j\right), & \text{$\gamma \geqslant N-1$}.\qquad \qquad \qquad \qquad \qquad \qquad \qquad
   \end{cases}\,
\end{multline}

With the equation (\ref{A1}) and the definition in equation (\ref{15}) and (\ref{17}), we can prove another important formula.

\begin{equation}\label{A2}
    \sum_{k=1}^N a(P_{kN})\left[P_{kN}\prod_{i>j}^{N-1}\left(x_i-x_j\right) \xi_{N-1}^l\right]\\
    =\begin{cases}
    0, & \text{$l<N-1$},\\
    \sum_{i=0}^{l-N+1}\frac{l!(-N)^{l-i}}{(N-1)^l i!(l-i)!}\sigma_1^iV_{(l-N+1-i)}, & \text{$l\geqslant N-1$}.
    \end{cases}
\end{equation}

$r_{_{N-1}}$ is invariant by exchanging the first $N-1$ particles, so we can divide all the exchange into two steps: First, exchange all the first $N-1$ particles; Second, exchange the last particle with the first $N-1$ particles.

\begin{multline}\label{A3}
    \qquad \qquad \qquad \psi(\lambda)=\sum_{k=1}^{N}\left\{a(P_{k,N})P_{k,N}\left[\sum_P a(P)P\varphi_1\left(\xi_1\right)
    \varphi_2\left(\xi_2\right)\cdots\varphi_{_{N-2}}\left(\xi_{_{N-2}}\right)\right]\varphi_{_\lambda}\left(\xi_{_{N-1}}\right)\right\}\\
    =C\sum_{k=1}^{N}\left[a(P_{k,N})P_{k,N}
    \prod_{i>j}^N\left(x_i-x_j\right)\exp \left(\sum_{i=1}^{N-1}-\frac{1}{2}\alpha_i^2\xi_i^2\right)\varphi_{_\lambda}\left(\xi_{_{N-1}}\right)\right]\qquad \quad\\
    =C\exp\left(-\frac{m\omega}{2\sqrt{N}\hbar}\sum_{i>j}^Nx_{ij}^2\right)\sum_{k=1}^{N}\left[a(P_{k,N})P_{k,N}\prod_{i>j}^{N-1}\left(x_i-x_j\right)H_\lambda\left(\alpha_{_{N-1}} \xi_{_{N-1}}\right)\right],\qquad \qquad \;
\end{multline}
with $H_\lambda\left(\alpha_{_{N-1}}\xi_{_{N-1}}\right)$ is the Hermite polynomial. Substitute equation (\ref{A2}) into equation (\ref{A3}), then it becomes:
When $\lambda<N-1$, $\psi(\lambda)=0$; When $\lambda\geqslant N-1$, define $\lambda'=\lambda-N+1$, then

\begin{equation}\label{A4}
  \psi(\lambda')=
  C\prod_{i>j}^N\left(x_i-x_j\right)\exp\left(-\frac{m\omega}{2\sqrt{N}\hbar}\sum_{i>j}^Nx_{ij}^2\right)\sum_{i=0}^{\frac{\lambda'-\epsilon(\lambda')}{2}}\left[\sum_{j=0}^{2i+\epsilon(\lambda')}
  \frac{(-1)^{i+j}\left(\frac{4N\sqrt{N}m\omega}{(N-1)\hbar}\right)^i\sigma_1^jV_{(2i-j+\epsilon(\lambda'))}}{(\frac{\lambda-\epsilon(\lambda')}{2}-i)!j!(N+2i-j-1+\epsilon(\lambda'))!N^j}\right].
\end{equation}
Next we give the simplified forms of $\psi(\lambda')$ when $\lambda'=N-1,\,\, N$ and $N+1$. two of which are nonzero, which may be useful to someone else.

\begin{equation*}
  \begin{split}
  &\psi(N-1)=C_0\prod_{i>j}^N\left(x_i-x_j\right)\exp\left(-\frac{m\omega}{2\sqrt{N}\hbar}\sum_{i>j}^Nx_{ij}^2\right), \,\,\,\,\,\,\,\,\,\,\,\,\, \psi(N)=0,\\
  &\psi(N+1)=C_1\left[\sum_{i>j}^N\left(x_i-x_j\right)^2-\frac{\hbar}{2m\omega}\sqrt{N}\left(N^2-1\right)\right]\prod_{i>j}^Nx_{ij}\exp\left(-\frac{m\omega}{2\sqrt{N}\hbar}\sum_{i>j}^Nx_{ij}^2\right).\\
  &C_0=\left[\left(\frac{m\omega}{\hbar}\right)^{\frac{N^2-1}{2}}\pi^{\frac{1-N}{2}}N^{\frac{N^2-3}{4}}\prod_{i=1}^N \frac{2^{i-1}}{i!}\right]^{1/2},\quad C_1=\left[\left(\frac{m\omega}{\hbar}\right)^{\frac{N^2+3}{2}}\frac{2}{N(N^2-1)}\pi^{\frac{1-N}{2}}N^{\frac{N^2-3}{4}}\prod_{i=1}^N \frac{2^{i-1}}{i!}\right]^{1/2}.
  \end{split}
\end{equation*}
So far, we have proved the correctness of equation (\ref{16}). When $\lambda\geqslant N$, define $k=\lambda-N=\lambda'-1$, rewrite equation (\ref{A4}), we can get equation (\ref{14}).

\end{document}